\documentclass[aps,pra,twocolumn,superscriptaddress,showkeys,showpacs]{revtex4-2}
\usepackage[colorlinks,linkcolor=blue,urlcolor=blue,anchorcolor=blue,citecolor=blue]{hyperref}
\usepackage{amsmath}
\usepackage{graphicx}
\usepackage{dcolumn}
\usepackage{mathrsfs}
\usepackage{amssymb}
\usepackage{amsfonts,multirow}
\usepackage{bm,lineno}
\usepackage{color}

\begin{document}
\title{Implementation and topological characterization of Weyl exceptional rings in quantum-mechanical systems}
\author{Hao-Long Zhang}\thanks{These authors contributed equally to this work.}
\affiliation{Fujian Key Laboratory of Quantum Information and Quantum Optics, College of Physics and Information Engineering, Fuzhou University, Fuzhou 350108, China}
\author{Pei-Rong Han}\thanks{These authors contributed equally to this work.}
\affiliation{School of Physics and Mechanical and Electrical Engineering, Longyan University, Longyan 364012, China.}
\affiliation{Fujian Key Laboratory of Quantum Information and Quantum Optics, College of Physics and Information Engineering, Fuzhou University, Fuzhou 350108, China}
\author{Xue-Jia Yu}\thanks{These authors contributed equally to this work.}
\author{Shou-Bang Yang}
\author{Jia-Hao L\"{u}}
\author{\\ Wen Ning}\email{ningw@fzu.edu.cn}
\author{Fan Wu}
\affiliation{Fujian Key Laboratory of Quantum Information and Quantum Optics, College of Physics and Information Engineering, Fuzhou University, Fuzhou 350108, China}
\author{Qi-Ping Su}
\author{Chui-Ping Yang}
\affiliation{School of Physics, Hangzhou Normal University, Hangzhou 311121, China}
\author{Zhen-Biao Yang}\email{zbyang@fzu.edu.cn}
\author{Shi-Biao Zheng}\email{t96034@fzu.edu.cn}
\affiliation{Fujian Key Laboratory of Quantum Information and Quantum Optics, College of Physics and Information Engineering, Fuzhou University, Fuzhou 350108, China}
\affiliation{Hefei National Laboratory, Hefei 230088, China}

\begin{abstract}
Non-Hermiticity can lead to the emergence of many intriguing phenomena that are absent in Hermitian systems, enabled by exceptional topological defects, among which Weyl exceptional rings (WER) are particularly interesting. The topology of a WER can be characterized by the quantized Berry phase and a nonzero Chern number, both encoded in the eigenvectors of the non-Hermitian Hamiltonian. So far, WERs have been realized with classical wave systems, whose eigenvectors can be well described by classical physics. We here report the first quantum-mechanical implementation of WERs and investigate the related topology transitions. The experiment system consists of a superconducting qubit and a dissipative resonator, coupled to each other. The high flexibility of the system enables us to characterize its eigenvectors on different manifolds of parameter space, each of which corresponds to a quantum-mechanical entangled state. We extract both the quantized Berry phase and Chern number from these eigenvectors, and demonstrate the topological transition triggered by shrinking the size of the manifold.
\end{abstract}

\maketitle

\vspace{0.5em}
\noindent\rule{\linewidth}{0.4pt}

\noindent\textbf{Received:} 26-Mar-2025 \quad
\textbf{Revised:} 28-Apr-2025 \quad
\textbf{Accepted:} 19-May-2025

\vspace{0.5em}
\noindent\rule{\linewidth}{0.4pt}

\section{Introduction}
Although most quantum-mechanical phenomena are observed by isolating the quantum systems from their surrounding environment so as to minimize the decoherence effects arising from interaction with the environment, the non-Hermitian (NH) effects due to dissipations can sometimes cause novel features that are inaccessible otherwise \cite{Phys.Rev.Lett.89.270401, Science.363.7709, Nat.Mater.18.783, Phys.Rev.Lett.131.260201}. The rich physics of non-Hermitian systems is closely associated with exceptional points (EPs), featuring the coalescence of both the eigenenergies and eigenstates. This enables EPs to display distinct properties compared to the degeneracies of Hermitian systems, where the eigenenergies coalesce but the eigenstates can remain orthogonal. Among these, exceptional topology is particularly appealing, which can manifest in either the eigenspectra or in the eigenvectors of the NH Hamiltonian \cite{Rev.Mod.Phys.93.015005, Nat.Rev.Phys.4.745}. The topological invariants of isolated EPs have been measured in different systems \cite{Science.359.1009, Sci.Adv.7.eabj8905, Phys.Rev.Lett.127.034301, Phys.Rev.Lett.127.090501, Phys.Rev.Lett.127.090501, Science.371.1240, Phys.Rev.Lett.130.017201, Phys.Rev.Lett.130.163001, Phys.Rev.A.108.052409,Nat.Commun.15.10293}.

The topology features associated with non-Hermiticity are further enriched by the discovery of one- and two-dimensional (2D) EP structures, such as EP
rings \cite{Phys.Rev.Lett.118.045701, Phys.Rev.B.99.121101, Phys.Rev.Lett.127.196801,Phys.Rev.B.104.L161117,Phys.Rev.Res.2.043268,Phys.Rev.B.100.245205,Nature.525.354,Nat.Photonics.13.623,Science.370.1077,Phys.Rev.Lett.129.084301, Nat.Commun.14.6660} and EP surfaces \cite{Phys.Rev.Lett.123.237202, Optica.6.190}. When some control parameter in the Hamiltonian is extended from the real domain to the complex domain, each EP pair is transformed into a ring, referred to as the Weyl exceptional ring (WER). It was discovered that a WER formed by second-order EPs (EP2s) carries a quantized Berry charge, that can be characterized by a Chern number obtained by integrating the Berry curvature over a closed 2D surface encompassing the ring, as well as by a quantized Berry phase associated with the integral of the Berry connection along a 2D loop encircling the ring \cite{Phys.Rev.B.99.121101}. WERs have been observed in several experiments \cite{Nature.525.354,Nat.Photonics.13.623,Science.370.1077,Phys.Rev.Lett.129.084301,Nat.Commun.14.6660}, but all are restricted to classical systems without any quantum effect. Even for the classical implementations, the associated Chern number has not been observed so far.

We here investigate the topological transition associated with the WER encoded in the entangled eigenvectors of a dissipative Jaynes-Cummings (JC) model. This model consists of a qubit and a decaying resonator, engineered with a circuit quantum electrodynamics (QED) architecture. One of the superconducting qubits in the circuit QED device is coupled to its readout resonator with an ac flux that produces a longitudinal parametric modulation to the qubit's transition frequency. The effective qubit-resonator coupling strength and detuning are tunable by the modulation amplitude and frequency. In the absence of dissipation, the Berry topological charge is carried by the degeneracy of the qubit-boson entangled eigenstates, which resides at the origin of the parameter space. The dissipation extends the point-like singularity to a ring, realizing a WER in the parameter space. The Berry curvature, which serves as a fictitious magnetic field, is extracted from the eigenvectors measured for different settings of the control parameter. By reducing the size of the manifold so as not to enclose the WER, the system undergoes a topological transition, manifested by an abrupt change of the Berry phase and the Chern number.
\begin{figure*}
    \centering
    \includegraphics{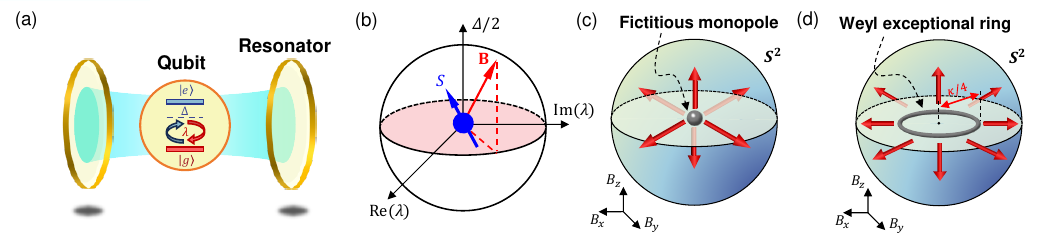}
    \caption{Construction of the WER. (a) The JC model. The system is composed of a photonic mode stored in a resonator interacting with a qubit, whose upper and lower levels are respectively denoted as $\left\vert e\right\rangle $ and $\left\vert g\right\rangle $. The qubit-resonator coupling coefficient and detuning are $\lambda $ and $\Delta $, respectively. (b) Spin representation. In the single-excitation subspace $\left\{ \left\vert \uparrow \right\rangle \equiv \left\vert e,0\right\rangle {\bf,}  \left\vert \downarrow \right\rangle \equiv \left\vert g,1\right\rangle \right\} $, the dynamics of the composite system is mathematically equivalent to a spin-1/2 ($S$) moving in a magnetic field ${\bf B}$, whose $x$-, $y$-, and $z$-components correspond to ${\rm  Re} (\lambda)$, ${\rm  Im}(\lambda)$, and $\Delta /2$, respectively. (c) Point-like topological defect. In the absence of dissipation, the eigenspectrum displays a two-fold degeneracy at the origin of the 3D parameter space $\left\{ B_{x}, B_{y}, B_{z}\right\}$, which can be considered as a fictitious monopole carrying a quantized topological charge. (d) WER. The non-Hermiticity is manifested by the photonic dissipation of the resonator with a rate $\kappa $. Due to the presence of dissipation, the point-like singularity is extended to a ring with the radius of $B_{WER}=\kappa /4$, centered at the origin and located on the $xy$-plane of the parameter space.}
    \label{fig1}
\end{figure*}

\section{Model and Methods}
We first consider the topology encoded in the parameter space of the JC model without dissipation (Fig.~\ref{fig1}a). In the framework rotating at the frequency of the resonator, the system Hamiltonian is given by (setting $\hbar = 1$) \cite{Proc.IEEE.51.89,J.Mod.Opt.40.1195}
\begin{equation}
\label{Hs}
H_{S}=\Delta \left\vert e\right\rangle \left\langle e\right\vert +\lambda
a^{\dagger }\left\vert g\right\rangle \left\langle e\right\vert +\lambda
^{\ast }a\left\vert e\right\rangle \left\langle g\right\vert ,
\end{equation}%
where $\left\vert e\right\rangle $ and $\left\vert g\right\rangle $ are the upper and lower levels of the qubit, $a^{\dagger }$ and $a$ are the creation and annihilation operators for the quantized field stored in the resonator, and $\lambda $ and $\Delta $ denote the qubit-cavity coupling coefficient and detuning, respectively. In the single-excitation subspace, the composite qubit-resonator system can be taken as a spin, with the basis states $\left\vert e,0\right\rangle $ and $\left\vert g,1\right\rangle $ respectively corresponding to the spin-up and -down states $\left\vert \uparrow \right\rangle $ and $\left\vert \downarrow \right\rangle $, where the number in each ket denotes the photon number of the resonator. With this analogy, the system dynamics can be described as the motion of the spin in a magnetic field ${\bf B}$ with the components $B_{x}=\mathop{\rm Re}(\lambda) $, $B_{y}=\mathop{\rm Im}(\lambda) $, and $B_{z}=\Delta /2$, as shown in Fig.~\ref{fig1}b. This Hermitian system has a degeneracy at the origin, referred to as a diabolic point, where the two eigenenergies coalesce but the eigenvectors do not. This degeneracy is a mathematic analog of the magnetic monopole that carries a quantized topological charge in the parameter space, as shown in Fig.~\ref{fig1}c. Due to the presence of such a topological defect, any manifold that encloses this singularity in the parameter space is topologically distinct from those without involving it. The fictitious magnetic field emanated from the topological charge is manifested by the Berry curvature. The quantized Berry flux penetrating through the manifold is characterized by the Chern number, defined as the integral of the Berry curvature over the manifold.
\begin{figure*}
    \centering
    \includegraphics[width=1\textwidth]{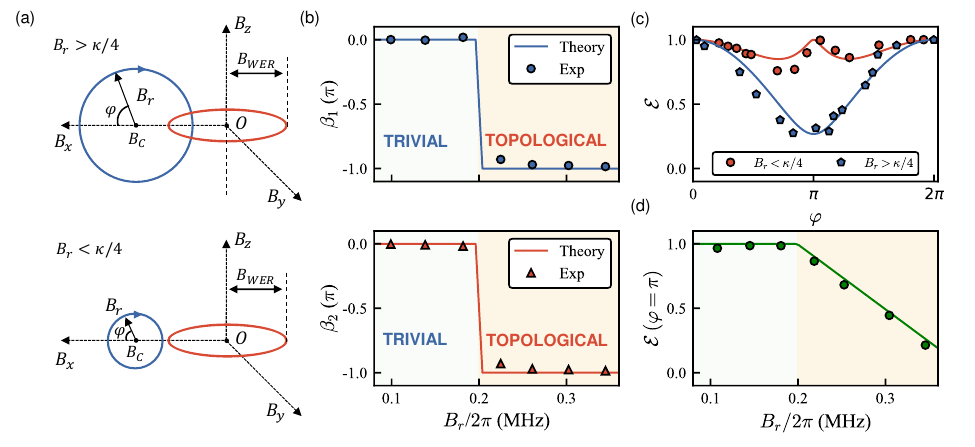}
    \caption{Topological transition characterized with the Berry phase. (a) Loops associated with topologically distinct phases. The loops are circularly-shaped, and centered at ($B_x = \kappa /2, B_z = 0$) on the $B_{x}$-$B_{z}$ plane. The WER is encircled when the radius of the traversed loop ($B_{r}$) is in the range $(\kappa /4,3\kappa /4)$. (b) Measured Berry phase versus $B_{r}$. The dots and triangles denote the results associated with the two pairs of eigenvectors $\left\{ \left\vert u_{n}^{r}({\bf B})\right\rangle,\left\vert u_{n}^{l}({\bf B})\right\rangle \right\} $ with $n=1,2$. (c) Measured qubit-resonator concurrence (${\cal E}$) versus $\varphi $ for different values of $B_{r}$ ($B_r/2\pi=0.18$, $0.34$ MHz). At each point, the concurrences are obtained from the corresponding right eigenvectors. (d) $ {\cal E}(\pi )$ versus $B_{r}$.} 
    
    
    \label{fig2}
\end{figure*}

With the dissipation of the photonic field being included, the system dynamics is described by a master equation. Conditional upon no photon being leaked into the environment, the system dynamics is described by the NH Hamiltonian \cite{J.Phys.A.40.14527, Int.J.Geom.Methos.Mod.Phys.8.1799, Phys.Rev.A.105.052208, Phys.Rev.A.105.022218} 
\begin{equation}\label{eqs2}
H_{NH}=H_{S}-\frac{1}{2}i\kappa a^{\dagger }a,
\end{equation}%
with $\kappa$ being the decaying rate of the photonic field. The eigenvectors of the NH Hamiltonian are drastically different from those of its Hermitian counterpart. Due to the non-Hermiticity, the left and right eigenvectors, defined as $H_{NH}\left\vert u_{n}^{r}\right\rangle =E_{n}\left\vert u_{n}^{r}\right\rangle $ and $\left\langle u_{n}^{l}\right\vert H_{NH}=\left\langle u_{n}^{l}\right\vert E_{n}$, are not the Hermitian conjugates of each other, and need to be obtained separately. For the present NH Hamiltonian, the two non-orthogonal right eigenvectors, $\vert u_n^r\rangle$ ($n=1,2$), in the single-excitation subspace can be written as
\begin{equation}
\left\vert u_{n}^{r}\right\rangle =\frac{\lambda ^{\ast }\left\vert
e,0\right\rangle +(E_{n}-\Delta )\left\vert g,1\right\rangle }{\sqrt{%
\left\vert \lambda \right\vert ^{2}+\left\vert E_{n}-\Delta \right\vert ^{2}}%
},
\end{equation}%
where 
\begin{equation}
E_{1,2}=\frac{2\Delta -i\kappa }{4}\pm \sqrt{\left\vert \lambda \right\vert
^{2}+\frac{(2\Delta +i\kappa )^{2}}{16}}.
\end{equation}%

The left eigenvectors can be obtained from the biorthonormal condition \cite{Rev.Mod.Phys.93.015005, Nat.Rev.Phys.4.745}
\begin{equation}
\left\langle u_{n}^{l}\right\vert \left. u_{m}^{r}\right\rangle =\delta
_{m,n}.
\end{equation}%

The non-Hermiticity changes the point-like degeneracy into a WER with the radius of $B_{WER}=\kappa /4$, centered at the origin and located on the $B_x$-$B_y$ plane of the parameter space (Fig.~\ref{fig1}d). Along the WER, both the eigenenergies and eigenvectors coalesce. Due to the ring-like structure of the topological defect, the topology of the parameter-space manifold depends upon its position, as well as upon its size, which is fundamentally distinct from that of a Hermitian system.

\section{Results}
We engineer the WER and demonstrate its topological feature using a circuit QED device with a bus resonator ($R_{b}$) and 5 frequency-tunable qubits, one of which ($Q$), together with its readout resonator ($R$), is used to realize the spin-boson model. $Q$ has an energy relaxation time $T_{1}\approx 14.3\ \mu$s and a pure Gaussian dephasing time $T_{2}^{\ast }\approx 5.3\ \mu$s at its idle frequency $\omega _{I}/2\pi=6.0$ GHz, where it is transformed from the initial ground state $\left\vert g\right\rangle $ to the excited state $\left\vert e\right\rangle $ with a $\pi $ pulse. During application of the pulse, $Q$ is highly detuned and thus effectively decoupled from both $R_{b}$ and $R$ with frequencies $\omega _{b}/2\pi=5.584$ GHz and $\omega/2\pi =6.656$ GHz, respectively. The pulse sequence is sketched in the Supplemental Material. After this transformation, $Q$ is coupled to $R$ through application of an ac flux, modulating the qubit frequency as $\omega _{q}=\omega _{0}+\varepsilon \cos (\nu t)$ \cite{Phys.Rev.Lett.131.260201}, where $\omega _{0}$ is the mean frequency and $\varepsilon $ ($\nu$) is the modulation amplitude (frequency), respectively. The modulation frequency $\nu \approx 2\pi \times$ ($678.4\sim 643.0$) MHz is close to $\omega -\omega _{0}$, so that $Q$ is quasi-resonantly coupled to $R$ at the first upper sideband of the modulation. The couplings at the carrier and other sidebands can be discarded due to large detunings. The system evolution is approximately governed by the Hamiltonian of Eq.~(\ref{Hs}) with the effective coupling $\lambda =\lambda _{r}J_{1}(\mu )$ and the detuning $\Delta =\nu +\omega _{0}-\omega $, where $\mu =\varepsilon /\nu $, $J_{1}(\mu )$ is the first-order Bessel function of the first kind, and $\lambda _{r}=2\pi\times 41$ MHz denotes the on-resonance coupling strength between the qubit and the readout resonator. The Hamiltonian parameters $\Delta $ and $\lambda $ are tunable by the parametric modulation pulse, and the non-Hermiticity is manifested by the photonic dissipation with a rate of $\kappa =5$ MHz. The decaying rates of $Q$ and $R_{b}$, $0.07$ MHz and $0.083$ MHz, are much smaller than $\kappa$, and thus can be discarded.  As R is initially in the vacuum state, the Q-R system evolves in the single-excitation subspace, so that both the qubit and the resonator cannot be coupled to energy levels with more than one excitation during their interaction.
\begin{figure*}
    \centering
    \includegraphics{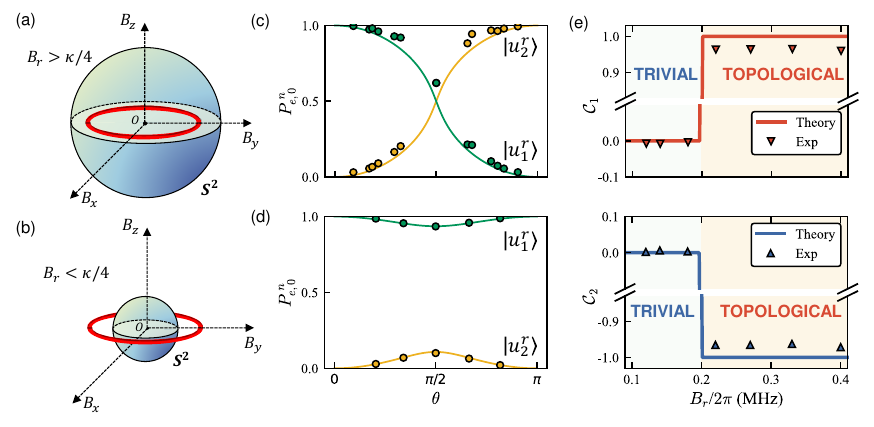}
    \caption{Topological transition characterized with Chern number. (a), (b) Manifolds associated with topologically distinct phases. The spherical manifolds are centered at the origin of the 3D parameter space. When the WER is enclosed in the manifold (a), the quantized Berry flux emanated from the WER pierces through the manifold. When the sphere is located inside the ring (b), the amount of the Berry flux entering the manifold equals the amount going out of it. (c), (d) Measured populations of $\left\vert e,0\right\rangle $ in $\left\vert u_{n}^{r}(\theta ,\phi )\right\rangle $ ($n=1,2$) versus $\theta$ for $B_{r}/2\pi=0.33$ MHz (c) and $0.14$ MHz (d). Here ($B_{r}, \theta, \phi$) denote the spherical coordinates for the control parameter. All data are measured for $\phi =0$. The lines denote the functions fitted with the measured data. (e) Measured Chern number versus the radius of the manifold. The dots and triangles denote the results associated with the two right eigenvectors $\left\vert u_{1}^{r}({\bf B})\right\rangle $ and $\left\vert u_{2}^{r}({\bf B})\right\rangle $, respectively.}
    \label{fig3}
\end{figure*}

The topological features of the engineered WER are manifested by the quantized Berry phase of a loop encircling the WER, as well as by the Chern number of a manifold enclosing it. The Berry phase is defined as \cite{Phys.Rev.Lett.118.045701}
\begin{equation}
\beta _{n}=i%
\displaystyle\oint %
\limits_{2{\cal L}}\left\langle u_{n}^{l}({\bf B})\right\vert \frac{\partial 
}{\partial {\bf B}}\left\vert u_{n}^{r}({\bf B})\right\rangle \cdot {{\rm d}\bf B,}
\end{equation}%
where the path $2{\cal L}$ travels across the ring twice along the loop in parameter space (Fig.~\ref{fig2}a), so that the eigenvector returns to the original one after the entire trajectory. When the loop encircles the WER, the acquired Berry phase is $\pm \pi$, which becomes zero for a loop without encircling it. Due to the chiral nature associated with the non-Hermiticity, the system does not adiabatically follow a specific eigenvector by slowly changing the control parameter \cite{Phys.Rev.X.8.021066}. The breakdown of the adiabaticity prevents measurement of the Berry phase by adiabatic evolution. However, the Berry phase associated with each eigenvector is encoded in the parameter space, and has no direct relation to the evolution time. This enables us to infer the Berry phase by measuring the dependence of the eigenvector on the control parameter \cite{Phys.Rev.Lett.131.260201}, without resorting to the adiabatic process.

The eigenvectors for a preset control parameter can be extracted from the joint $Q$-$R$ output states dynamically evolved for different times under the NH Hamiltonian of Eq.~(\ref{eqs2}). The interaction time is controlled by the parametric modulating pulse. When the modulation is switched off, $Q$ is effectively decoupled from $R$ due to large detuning. Then $Q$'s state is mapped to an ancilla qubit $Q_{a}$ with the assistance of $R_{b}$, following which $R$'s state is transferred to $Q$ through the modulation-induced sideband interaction. The resulting joint $Q_{a}$-$Q$ state projected to the single-excitation subspace, which can be measured by the quantum state tomography and postselection technique \cite{Phys.Rev.Lett.131.260201}, corresponds to the $Q$-$R$ output state just before the state transfer procedure. The eigenvectors are inferred from the tracked time-evolving $Q$-$R$ output states, as detailed in the Supplemental Material. We choose circular loops centered at ($B_{z}=0$, $B_{x}=\kappa /2$) on the $B_{x}$-$B_{z}$ plane. With this choice, whether or not the WER is encircled depends on the radius of the traversed loop ($B_{r}$), as shown in Fig.~\ref{fig2}a. We can fit right eigenvectors $\left\vert u_{n}^{r}({\bf B})\right\rangle $ as functions of the control parameter ${\bf B}$ with the results extracted for different settings of ${\bf B}$, and calculate the left vectors $\left\vert u_{n}^{l}({\bf B})\right\rangle $ using the biorthonormal condition. Fig.~\ref{fig2}b shows the Berry phase ($\beta_1$ and $\beta_2$) for the two pairs of eigenvectors versus $B_{r}$. These Berry phases, measured at $B_{r}=0.427\kappa$, are about $-0.9844\pi$ and $-0.9844\pi$, respectively. With the shrinking of the loop, each of these Berry phases makes an abrupt change around $B_{r}=0.226\kappa$, quickly dropping to $0$ when crossing this critical point, thereby manifesting a topological transition. Owing to control errors, such an experimentally inferred critical point slightly deviates from the theoretical value $\kappa/4$. We note that it is experimentally challenging to extract the eigenstates when the control parameter is infinitely close to the critical point. This is due to the fact that both the eigenstates and eigenenergies tends to coalesce, so that the state evolution speeds become infinitely slow in this case. Consequently, the jump of the Berry phase from $-\pi$ to $0$ at the critical point cannot be unambiguously confirmed. In our experiment, the minimum (maximal) value of $B_r$ for the observed trivial (topological) phase is $0.126 \kappa$ ($0.427\kappa$).

Unlike previous implementations of WERs \cite{Rev.Mod.Phys.93.015005,Nat.Rev.Phys.4.745,Nature.525.354,Nat.Photonics.13.623,Phys.Rev.Lett.129.084301}, in our system the topological characteristic is encoded in highly nonclassical states of the joint $Q$-$R$ system. In the single-excitation subspace, the quantum entanglement between $Q$ and $R$ can be well measured by the concurrence ${\cal E}$ \cite{Phys.Rev.Lett.80.2245}. The concurrences of each right eigenvector versus $\varphi$ for different values of $B_{r}$ are presented in Fig.~\ref{fig2}c, where $\varphi$ is the angle between the control parameter ${\bf B}$ and the x-axis in the displaced frame, whose origin coincides with the center of the loops. The solid lines denote the results for the ideal eigenvectors. We note that the topological phase transition is closely related to the exceptional entanglement behavior. When $B_{r}<B_{C}-B_{WER}=\kappa /4$, the system works above the EP where $B_{x}=\kappa /4$ and $B_{y}=B_{z}=0$ for both $\varphi =0$ and $\pi$, and consequently ${\cal E}(0)={\cal E}(\pi )=1$ in theory \cite{Phys.Rev.Lett.131.260201}. Here $B_{C}=\kappa /2$ denotes the magnitude of ${\bf B}$ at the center of the loops. For $B_{r}>\kappa /4$, ${\cal E}(0)=1$, while ${\cal E}(\pi )$ linearly increases with $B_{C}-B_{r}$. This implies that ${\cal E}(\pi )$ is independent of $B_{r}$ for the trivial phase, but exhibits a $B_{r}$-dependence for the topological phase. To confirm this point, in Fig.~\ref{fig2}d, we display the measured ${\cal E}(\pi )$ versus $B_{r}$, which is well agreement with the ideal result (solid line). Consequently, the derivative of ${\cal E}(\pi )$ with respect to $B_{r}$ is not continuous at the point $B_{r}=\kappa /4$ \cite{Phys.Rev.Lett.131.260201}. These results clearly demonstrate that the geometric features of the system are encoded in the entangled states of $Q$ and $R$, illustrating the quantum-mechanical character of the measured Berry phase, and the topological phase transition coincides with the exceptional entanglement phase transition \cite{Phys.Rev.Lett.131.260201}.


It is convenient to calculate the Chern number on a spherical manifold, which is centered at the origin of the parameter space. For a sphere with a specific radius $B_{r}$, the control parameters are the polar and azimuth angles ($\theta,\phi $). The $\phi $ and $\theta $ components of the Berry connection are defined as
\begin{eqnarray}
A_{\phi }^{n} &=&i\left\langle u_{n}^{r}(\theta ,\phi )\right\vert \partial
_{\phi }\left\vert u_{n}^{r}(\theta ,\phi )\right\rangle , \\
A_{\theta }^{n} &=&i\left\langle u_{n}^{r}(\theta ,\phi )\right\vert
\partial _{\theta }\left\vert u_{n}^{r}(\theta ,\phi )\right\rangle . 
\nonumber
\end{eqnarray}%
The Chern number is given by
\begin{equation}
{\cal C}_{n}=\frac{1}{2\pi }%
\displaystyle\int %
\limits_{0}^{2\pi }{\rm d}\phi 
\displaystyle\int %
\limits_{0}^{\pi }{\rm d}\theta F_{\theta \phi }^{n},
\end{equation}%
where $F_{\theta \phi }^{n}=\partial _{\theta }A_{\phi }^{n}-\partial _{\phi}A_{\theta }^{n}$ denotes the Berry curvature \cite{Proc.R.Soc.Lond.A.392.45, Phys.Rev.Lett.118.045701,arxiv.2402.12755}. When $B_{r} > \kappa /4$, the WER is enclosed in the manifold (Fig.~\ref{fig3}a), so that the quantized Berry flux emanated from the WER pierces through the manifold, and the Chern number is $\pm 1$. For $B_{r}<\kappa /4$, the sphere is located inside the ring (Fig.~\ref{fig3}b), the Berry flux entering the manifold and that going out of it cancel out each other, resulting in a zero Chern number. 

Due to the spherical symmetry, the Berry connection $A_{\theta }^{n}$\ is independent of $\phi$, so that $F_{\theta \phi }^{n}=\partial _{\theta}A_{\phi }^{n}$. Taking advantage of this symmetry, the Chern number can be obtained from the results measured along $0^{\circ}$-meridian \cite{Phys.Rev.Lett.113.050402, Nature.515.237},
\begin{eqnarray}
    {\cal C}_{n}&=&\int_{0}^{\pi}{\rm d}\theta\partial_{\theta}A_{\phi}^n\\
    &=&P_{e,0}^n(\pi)-P_{e,0}^n(0), 
    \nonumber
\end{eqnarray}
where $P_{e,0}^n(\theta)$ is the population of $\vert e,0\rangle$ in the eigenvector $\vert u_{n}^{r}(\theta, 0)\rangle$. With data of $P^{n}_{e, 0}$ measured for different values of $\theta $, we can fit the function $P_{e,0}^{n}(\theta )$, which are shown in Fig.~\ref{fig3}c and d. In our experiment,  it is difficult to extract the eigenvectors when $\lambda$ is small (see Supplemental Material for details). Here we calculate the Chern number using the $P_{e,0}^n(\theta)$ with $\theta$ lying as close to $\pi$ and $0$ as possible. Fig.~\ref{fig3}e displays thus-obtained Chern numbers ${\cal C}_{1}$ and ${\cal C}_{2}$ versus the radius of the manifold. As expected, the Chern numbers, measured on a manifold with a radius larger than that of the WER $\kappa /4$, are equal to $\pm 1$. When $B_{r}$ is reduced to the critical value of $\kappa /4$, a topological transition occurs, characterized by an abrupt drop of the Chern number to $0$. These results can be interpreted as follows. Under the condition that the WER is enclosed in the manifold, the eigenvector $\left\vert u_{n}^{r}(\theta,0)\right\rangle $ makes a $\pi $ rotation, flipping from $\left\vert g,1\right\rangle $ ($\left\vert e,0\right\rangle $) to $\left\vert e,0\right\rangle $ ($\left\vert g,1\right\rangle $) following the variation of $\theta $ from $0$ to $\pi $, as illustrated in Fig.~\ref{fig3}c, where $B_{r}/2\pi=0.33$ MHz. In distinct contrast, for $B_{r}<\kappa /4$, $\left\vert u_{n}^{r}(\theta ,0)\right\rangle $ tilts away from vertical with an angle smaller than $\pi /2$ and then returns (Fig.~\ref{fig3}d, $B_{r}/2\pi=0.18$ MHz). The measured $P_{e,0}^{n}$ versus $\theta $ for different values of $B_{r}$ are detailed in Supplemental Material. We note this topological behavior represents a unique characteristic that distinguishes the NH system from Hermitian ones, where a topological transition occurs only when the manifold is displaced so as not to enclose the origin of parameter$-$the diabolic point \cite{Nat.Commun.15.10293, Phys.Rev.Lett.127.034301,Phys.Rev.Lett.113.050402,Nature.515.237,Phys.Rev.Lett.122.210401,Science.360.1429,Phys.Rev.Lett.126.017702,Science.375.1017,Nat.Sci.Rev.7.254}, but cannot be realized by shrinking the manifold centered at the diabolic point. Because of the experimental limitation, the minimum (maximal) value of $B_{r}$ for the observed trivial (topological) phase is $0.151\kappa$ ($0.503\kappa$).

\section{Discussion and conclusion}
In conclusion, we have constructed a WER in a system composed of a qubit coupled to a resonator supporting a decaying photonic field, and characterized its topological features by the quantized Berry phase and Chern number, which are inferred from the measured eigenvectors of the governing NH Hamiltonian. We have observed a topological phase transition featuring a change of the Chern numbers from $\pm 1$ to 0, by continually reducing the size of the spherical manifold centered at the origin of the parameter space. Our results confirm that the non-Hermiticity changes a point-like degeneracy into a 2D ring of EPs, giving arise to exceptional topological phenomena that are absent in Hermitian systems. Our method can be generalized to realize the EP ring associated with a non-Abelian monopole and to explore NH topological transitions based on a higher-order Chern number.

\section*{Conflict of interest}
The authors declare that they have no conflict of interest

\section*{Acknowledgments}
This work was supported by the National Natural Science Foundation of China (12274080, 12474356, 12475015, U21A20436) and Innovation Program for Quantum Science and Technology (2021ZD0300200, 2021ZD0301705).

\section*{Author Contributions}
Shi-Biao Zheng conceived the experiment. Wen Ning, Hao-long Zhang, and Pei-Rong Han supervised by Zhen-Biao Yang and Shi-Biao Zheng, carried out the experiment. Hao-long Zhang, Pei-Rong Han, and Xue-Jia Yu analyzed the data. Shi-Biao Zheng, Zhen-Biao Yang, and Xue-Jia Yu co-wrote the paper. Shou-Bang Yang, Jia-Hao Lü, Fan Wu, Qi-Ping Su, and Chui-Ping Yang helped to interpret the observed phenomena and to write the paper.

\end{document}


\title{Supplementary Material for \\ ``Observation of topological transitions associated with a Weyl
exceptional ring''}

\author{Hao-Long Zhang}\thanks{These authors contribute equally to this work.}
\affiliation{Fujian Key Laboratory of Quantum Information and Quantum Optics, College of Physics and Information Engineering, Fuzhou University, Fuzhou 350108, China}
\author{Pei-Rong Han}\thanks{These authors contribute equally to this work.}
\affiliation{School of Physics and Mechanical and Electrical Engineering, Longyan University, Longyan 364012, China.}
\affiliation{Fujian Key Laboratory of Quantum Information and Quantum Optics, College of Physics and Information Engineering, Fuzhou University, Fuzhou 350108, China}
\author{Xue-Jia Yu}\thanks{These authors contribute equally to this work.}
\author{Shou-Bang Yang}
\author{Jia-Hao L\"{u}}
\author{\\ Wen Ning}\email{ningw@fzu.edu.cn}
\author{Fan Wu}
\affiliation{Fujian Key Laboratory of Quantum Information and Quantum Optics, College of Physics and Information Engineering, Fuzhou University, Fuzhou 350108, China}
\author{Qi-Ping Su}
\author{Chui-Ping Yang}
\affiliation{School of Physics, Hangzhou Normal University, Hangzhou 311121, China}
\author{Zhen-Biao Yang}\email{zbyang@fzu.edu.cn}
\author{Shi-Biao Zheng}\email{t96034@fzu.edu.cn}
\affiliation{Fujian Key Laboratory of Quantum Information and Quantum Optics, College of Physics and Information Engineering, Fuzhou University, Fuzhou 350108, China}
\affiliation{Hefei National Laboratory, Hefei 230088, China}

\maketitle

\tableofcontents

\section{Engineering of the dissipative Jaynes-Cummings model}
\begin{figure}
    \centering
    \includegraphics{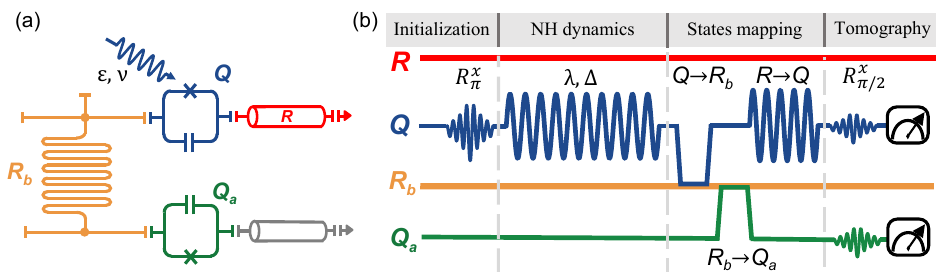}
    \caption{(online) (a) Implementation of the NH Hamiltonian. In the experimental system, a superconducting qubit $Q$ is highly detuned from the decaying resonator $R$, which has a fixed frequency $\omega_r/2\pi = 6.656$ GHz. The $Q$-$R$ interaction is enabled by an ac flux, which modulates $Q$'s energy gap around the mean value  $\omega_0$ with a frequency $\nu$. This modulation enables a photonic swapping coupling at one sideband, with the coupling strength controlled by the modulating amplitude $\varepsilon$. (b) Pulse sequence. The NH dynamics are initiated with the state $\vert e, 0\rangle$ prepared by performing a $\pi$ pulse on $\vert g, 0\rangle$. After the NH dynamics process, we perform the state mappings $Q\rightarrow R_b$, $R_b\rightarrow Q_a$, and $R\rightarrow Q$, each realized by an on-resonance swapping gate. Here, $Q_a$ is an ancilla qubit, and $R_b$ represents the bus resonator coupled to both qubits. Through this method, the $Q$-$R$ output state produced by the NH dynamics is encoded in the joint $Q_a$-$Q$ state, which can be measured by quantum state tomography.}
    \label{figS1}
\end{figure}

The dissipative Jaynes-Cummings (JC) model is engineered in a circuit quantum electrodynamics architecture, where a bus resonator ($R_b$) is connected to 5 Xmon qubits, each of which is individually connected to a readout resonator. The schematic diagram of the device is shown in Fig.~\ref{figS1}a. One of these qubits ($Q$) and its readout resonator ($R$) are used to construct the JC model. Another qubit ($Q_{a}$) is used as an ancilla for reading out  $R$'s state. The photonic swapping between $Q$ and $R$ is mediated by applying an ac flux to $Q$, modulating its frequency as
\begin{equation}\label{eqs1}
\omega _{q}=\omega _{0}+\varepsilon \cos (\nu t),
\end{equation}
where $\omega _{0}$ is  the mean frequency and $\varepsilon $ ($\nu$) is the modulation amplitude (frequency). The Hamiltonian of the $Q$-$R$ system is given by (setting $\hbar = 1$)

\begin{equation}
H=\omega _{r}a^{\dagger }a+\omega _{q}\left\vert e\right\rangle \left\langle
e\right\vert +\left(\lambda _{r}a^{\dagger }\left\vert g\right\rangle
\left\langle e\right\vert +H.c.\right),
\end{equation}
where $a^{\dagger}$ and $a$ denote the creation and annihilation operators for the photonic field stored in $R$, $\left\vert e\right\rangle$ and $\left\vert g\right\rangle$ are the excited and ground states of $Q$, and $\lambda _{r}$ is the on-resonance $Q$-$R$ coupling strength. We here have assumed that $Q$ is highly detuned from $R_b$ and the other qubits, so that the $Q$-$R$ dynamics are unaffected by them.

Under the condition $\left\vert \Delta \right\vert \ll \lambda _{r}$ where $\Delta =\omega _{0}+\nu -\omega _{r}$, $Q$ swaps photons with R at the first upper sideband with respect to the modulation. To derive the effective Hamiltonian, we perform the transformation $e^{i\int_{0}^{t}H_{0}dt}$, where $H_{0}=(\omega _{q}-\Delta )\left\vert e\right\rangle \left\langle e\right\vert +\omega _{r}a^{\dagger }a$. Then the Hamiltonian is
\begin{equation}\label{eqs3}
H_{S}=\Delta \left\vert e\right\rangle \left\langle e\right\vert +\left[e^{-i\mu
\sin (\nu t)}e^{i\delta t}\lambda _{r}a^{\dagger }\left\vert
g\right\rangle \left\langle e\right\vert +H.c.\right],
\end{equation}
where $\mu =\varepsilon /\nu $ and $\delta =\omega _{r}-(\omega _{0}-\Delta )$. With the application of the Jacobi-Anger expansion $e^{-i\mu \sin (\nu t)}=\stackrel{\infty }{\mathrel{\mathop{\sum }\limits_{n=-\infty }}}J_{n}(\mu )e^{-in\nu t}$, the Hamiltonian (\ref{eqs3}) can be expressed as
\begin{equation}\label{eqs4}
H_{S}=\Delta \left\vert e\right\rangle \left\langle e\right\vert +\stackrel{\infty }{\mathrel{\mathop{\sum }\limits_{n=-\infty }}}\left[J_{n}(\mu )e^{-in\nu t}e^{i\delta t}\lambda _{r}a^{\dagger }\left\vert g\right\rangle \left\langle e\right\vert +H.c.\right],
\end{equation}
where $J_{n}(\mu)$ is the $n$th Bessel function of the first kind. With the choice $\delta =\nu $ and for $\lambda _{r}\ll \nu $, the terms (with $n\neq 1$) oscillating fast can be discarded. Then the Hamiltonian (\ref{eqs4}) reduces to
\begin{equation}\label{eqs5}
H_{S}=\Delta \left\vert e\right\rangle \left\langle e\right\vert +\left(\lambda
a^{\dagger }\left\vert g\right\rangle \left\langle e\right\vert +H.c.\right),
\end{equation}
where $\lambda =\lambda _{r}J_{1}(\mu )$.

Suppose that the energy dissipation rate of $Q$ is much smaller than the photonic decaying rate of $R$, and thus can be neglected. Under the competition between the coherent dynamics and the photonic dissipation of $R$, the system evolution is described by the master equation
\begin{equation}
\frac{d\rho }{dt}=-i[H_{NH},\rho ]+\kappa a\rho a^{\dagger },
\end{equation}
where
\begin{equation}
H_{NH}=H_{S}-\frac{i\kappa }{2}a^{\dagger }a.
\end{equation}
For the initial state $\left\vert e,0\right\rangle $, the system output state is a mixture of the $Q$-$R$ entangled state in the subspace $\left\{\left\vert e,0\right\rangle ,\left\vert g,1\right\rangle \right\} $ and the state $\left\vert g,0\right\rangle $, which is produced by the photon-loss-induced quantum jump. The quantum state evolution trajectory without any quantum jump is governed by the NH Hamiltonian $H_{NH}$.

\section{System parameters and pulse sequence}\label{S2}
$R$ has a fixed frequency $\omega _{r}/2\pi=6.656$ GHz and a photonic decaying rate $\kappa =5$ MHz, which is much larger than $Q$'s energy relaxation rate $\gamma =0.07$ MHz. The frequency and photonic decaying rate of $R_b$ are $\omega _{b}/2\pi=5.584$ GHz and $\kappa =0.083$ MHz, respectively. Before the experiment, $Q$ is initialized to the ground state at its idle frequency $ \omega _{I}/2\pi=6.00$ GHz, where $Q$ is effectively decoupled from $R$ ($R_b$) as the detuning $2\pi\times 656$ MHz ($2\pi\times 416$ MHz) is much larger than the on-resonance $Q$-$R$ ($Q$-$R_b$) coupling strength, which is $2\pi \times 41$ MHz ($2\pi \times 20.9$) MHz.

The experiment involves three parts: preparation of the initial state, controlled $Q$-$R$ NH dynamics, and output state tomography, as sketched in Fig.~\ref{figS1}b. The initial state $\left\vert e,0\right\rangle $ is prepared by applying a $\pi$ pulse to $Q$. This is followed by the application of the parametric modulation to $Q$, mediating a sideband interaction between $Q$ and $R$ to realize the effective Hamiltonian $H_{S}$ of Eq. (\ref{eqs5}), with the parameters $\Delta $ and $\lambda$ controllable by the modulation frequency $\nu $ and amplitude $\varepsilon $. During this sideband interaction, all the other qubits stay at their idle frequencies. The detuning between their idle frequencies and the frequency of Q range from $2\pi \times 1.2$ GHz to $2\pi \times 1.4$ GHz, much larger than their couplings with $Q$ mediated by $R_b$, which are on the order of 1 MHz. These large detunings effectively decouple these qubits from $Q$ during the $Q$-$R$ sideband interaction.

After a preset interaction time, the modulation pulse is switched off, following which a joint state tomography is performed. However, $R$'s quantum state cannot be directly read out. To overcome this problem, we transfer $Q$'s state to $R_{b}$ by a swapping operation, realized by tuning $Q$ and $R_b$ on resonance for a time $\pi /(2g)$, where $g$ is their coupling strength. Following this, the state of $R_{b}$ is mapped to an ancilla qubit $Q_{a}$. Then $R$'s state is transferred to $Q$ through the first sideband interaction. The resulting joint $Q_{a}$-$Q$ state, corresponding to $Q$-$R$ state before the state transfer procedure, can be measured by quantum state tomography.

\section{Realization of longitudinal modulation and detuning of qubits frequency}\label{S3}
In superconducting quantum platforms, the qubit transition frequency is modulated by an ac flux, generated by applying a pulse to the Z line. In our superconducting sample, the qubit transition frequency needs to be periodically modulated at a frequency $\nu/2\pi > 0.65$ GHz, as specified in Eq.~(\ref{eqs1}). If the ac flux is modulated away from the sweet spot $\omega_m$ \cite{Appl.Phys.Rev.6.021318}, the modulation frequency $\nu$ of the qubit equals the frequency of the ac flux, which does not meet the experimental requirement given the limitations of our current hardware. To address this issue, we bias the flux to the sweet spot before applying longitudinal modulation. The modulation frequency of the qubit, achieved by modulating the flux at the sweet spot, is twice as high as that away from the sweet spot, enabling us to reach the desired modulation frequency. However, this method imposes an additional constraint, $\omega_r - \omega_0 = \omega_m - \varepsilon$, which restricts us in controlling the detuning $\Delta$ between $Q$ and $R$  by adjusting the mean frequency of the qubit.
\begin{figure}
    \centering
    \includegraphics{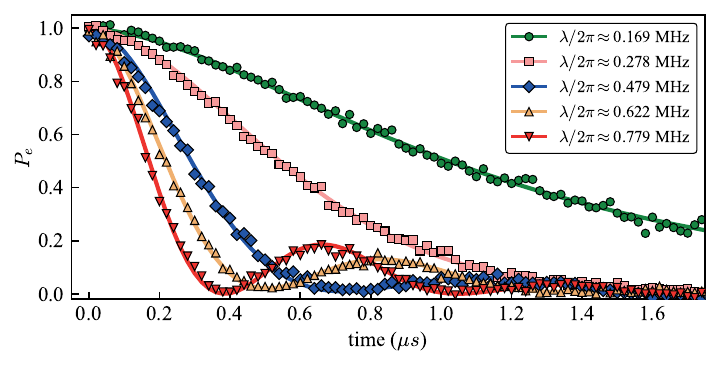}
    \caption{(online) Temporal evolution of the population $P_e$ under different longitudinal modulations.  The lines denote the fitting results and the dots represent the experimental results.}
    \label{fig_lm}
\end{figure}

For specific values of $\Delta$ and $\lambda$, we need to determine the corresponding values of modulation parameters to realize the effective Hamiltonian $H_S$. Initially, we need to determine the longitudinal modulation frequency $\nu$ and amplitude $\varepsilon$ that correspond to the effective coupling strength $\lambda$. Disregarding the detuning $\Delta$ ($\Delta=0$), the coherent dynamics of the system comprising $Q$ and $R$ is governed by Eq.~(\ref{eqs5}). Due to the additional constraints mentioned above, the modulation frequency $\nu$, the amplitude $\varepsilon$, and the mean frequency $\omega_0$ are deterministically related to the effective coupling strength $\lambda$ in a one-to-one manner. 

To determine this relationship, we first prepare $Q$ in the excited state $\vert e \rangle $, and then apply a longitudinal modulation while keeping $\varepsilon$ fixed. We observe oscillatory behavior in the population of the state $\vert e \rangle$ of the qubit over time for various modulation frequencies $\nu$. By identifying the point where the exchange between $Q$ and $R$ is almost complete, such set of $\nu$ and $\varepsilon$ determines an effective coupling between $Q$ and $R$. Subsequently, the effective coupling strength $\lambda$ under this parameter configuration is obtained by fitting the population of the state $\vert e \rangle$ using Eq.~(\ref{eqs5}) with $\Delta=0$. This process is repeated with finite-tuning $\nu$ and $\varepsilon$ until all necessary values for $\lambda$ are obtained. To validate these modulations, we present the vacuum Rabi oscillation signals for the test qubit $Q$ induced by these sideband couplings for different modulating amplitudes in Fig.~\ref{fig_lm}. Due to imperfections in the longitudinal modulation pulse, directly adjusting the mean frequency of the qubits by changing $\varepsilon$ becomes challenging. Instead, we achieve detuning by fine-tuning the frequency $\nu$ of the periodic modulation after observing first-order sideband resonances between $Q$ and $R$. At the point, it has $\nu=\omega_r - \omega_0$, $\mu = \varepsilon/\nu$ and $\lambda = \lambda_{r}J_1({\mu})$. Then the Hamiltonian is
\begin{equation}
H=e^{-i\mu^\prime\sin \left[\left(\nu + \Delta \right) t\right]}e^{i\left(\omega_r - \omega_0 \right) t}\lambda _{r}a^{\dagger }\left\vert g\right\rangle \left\langle e\right\vert +H.c.,
\end{equation}%
where $\mu^\prime=\varepsilon/(\nu + \Delta)$. By use of the Jacobi-Anger expansion, the Hamiltonian becomes
\begin{equation}
H=\stackrel{\infty }{\mathrel{\mathop{\sum }\limits_{n=-\infty }}}J_{n} \left(\mu^\prime \right)e^{-in(\nu+\Delta) t}e^{i \left(\omega_r - \omega_0 \right) t}\lambda _{r}a^{\dagger }\left\vert g\right\rangle \left\langle e\right\vert +H.c. .
\end{equation}%
The terms (with $n\neq 1$) oscillating fast can be discarded, thus $H$ reduces to
\begin{equation}
H_{S}=\Delta \left\vert e\right\rangle \left\langle e\right\vert +\left(\lambda^\prime
a^{\dagger }\left\vert g\right\rangle \left\langle e\right\vert +H.c. \right),
\end{equation}
where $\lambda^\prime = \lambda_{r}J_1(\mu^\prime)\ne \lambda$.
The effective coupling $\lambda$, in this case, deviates slightly from our preset value due to the imperfections of the modulated pulses, e.g., when we adjust the modulation frequency of qubit $\nu$, the corresponding $\varepsilon$  for the same pulse amplitude also experiences slight changes. Therefore, to obtain more accurate values for $\Delta$ and $\lambda$, we perform a preliminary fitting of the experimental data.

\section{Correction for state mapping}

As the excitation number is conserved under the NH Hamiltonian $H_{NH}$, for the initial state $\left\vert e,0\right\rangle $, we can postselect the output state without undergoing any quantum jump by discarding the measurement output, where both $Q$ and $Q_{a}$ are in their ground states. However, due to the strong dissipation of $R$, the $Q_{a}$-$Q$ state after the state mapping does not perfectly coincide with the $Q$-$R$ state just before the mapping.
\begin{figure}
    \centering
    \includegraphics{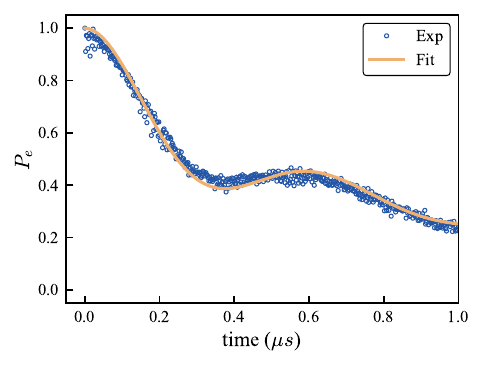}
    \caption{(online) The fitting results of calibration to parameters $\Delta$ and $\lambda$. The orange lines denote the fitting results and the dots represent the experimental results.}
    \label{figS5_1}
\end{figure}

Suppose that the $Q$-$R$ output state, associated with the no-jump trajectory governed by the NH Hamiltonian $H_{NH}$, is
\begin{equation}
\left\vert \psi _{NH}\right\rangle =\alpha \left\vert e,0\right\rangle
+\beta \left\vert g,1\right\rangle .
\end{equation}%
After the state mapping, the $Q_{a}$-$Q$ output state is
\begin{equation}
\left\vert \psi _{NH}^{\prime }\right\rangle =\alpha ^{\prime }\left\vert
e_{a},g\right\rangle +\beta ^{\prime }\left\vert g_{a},e\right\rangle ,
\end{equation}
where
\begin{eqnarray}
\alpha ^{\prime } &=&\frac{\alpha }{\sqrt{\left\vert \alpha \right\vert
^{2}+\beta e^{-\kappa (t_{1}+t_{2}/2)}}}, \\
\beta ^{\prime } &=&\frac{\beta e^{-\kappa (t_{1}/2+t_{2}/4)}}{\sqrt{%
\left\vert \alpha \right\vert ^{2}+\beta e^{-\kappa (t_{1}+t_{2}/2)}}},
\end{eqnarray}
with $t_{1}$ being the duration for the $Q\rightarrow R_{b}\rightarrow Q_{a}$ state mapping, and $t_{2}$ denoting the duration for the $R\rightarrow Q$ mapping. Given that  $t_1 \ll t_2$, we only consider $t_2$ and set $t_2=118$ ns in our experiment. We here neglect the dissipations of $Q$, $R_{b}$, and $Q_{a}$. This implies that we can infer the coefficients $\alpha$ and $\beta$ from the measured values of $\alpha ^{\prime}$ and $\beta ^{\prime}$.
\begin{figure}
    \centering
    \includegraphics{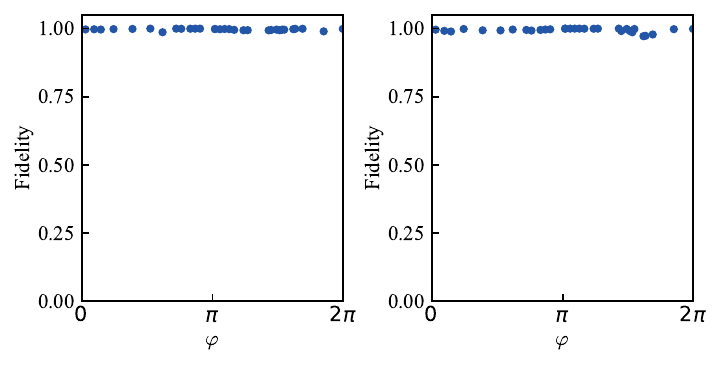}
    \caption{(online) Fidelities of fitted eigenvectors. (a) $\mathcal{F}_0$ as a function of $\varphi$. (b) $\mathcal{F}_1$ as a function of $\varphi$.}
    \label{figS5_2}
\end{figure}

\section{ Measurement of eigenvectors}
\label{S5}
As mentioned in Sec.~\ref{S3}, due to the experimental imperfections, the parameters $\Delta$ and $\lambda$ may have minor errors compared to the theoretical predictions obtained by the experimental parameters $\nu$ and $\varepsilon$. To calibrate these errors, we prepare the initial state $\vert e, 0\rangle$ and measure the evolution of the population of $\vert e\rangle$ with the predicted parameters $\lambda_i$ and $\Delta_i$. We then fit this population to obtain the calibrated parameters $\lambda_c$ and $\Delta_c$ based on the error function
\begin{equation}
    {\rm F^{1}_{err}} = \sum_{i=1}^{N}\left[P^{\rm exp}_{e,i}(t_i)-P^{\rm the}_{e,i}(t_i) \right]^2/N,
\end{equation}
where $P^{\rm exp}_{e, i}(t_i)$ is the evolution of the population of $\vert e\rangle$ at the moment $t_i$ and $P^{\rm the}_{e,i}(t_i)$ is the corresponding theoretical result. Based on this error function, the next step is to seek the fitting parameters to minimize the value of the error function. This allows us to obtain the calibrated parameters $\lambda_c$ and $\Delta_c$. The fitting results are shown in Fig.~\ref{figS5_1}, demonstrating the validity of this method.

After the calibration of the parameters $\Delta$ and $\lambda$, we can extract the corresponding eigenvectors of this system. The final state $\left\vert \psi _{NH}\right\rangle $ can be written as
\begin{equation}
\left\vert \psi _{NH}\right\rangle ={\cal N}\left(
C_{1}e^{-iE_{1}t}\left\vert u_{1}^{r}\right\rangle
-C_{2}e^{-iE_{2}t}\left\vert u_{2}^{r}\right\rangle \right) ,
\end{equation}%
where $\mathcal{N}$ is the normalization coefficient, the coefficients $C_{n}$ ($n=1,2$) are given by
\begin{equation}
C_{n}=\frac{\sqrt{\left\vert \lambda \right\vert ^{2}+\left\vert
E_{n}-\Delta \right\vert ^{2}}}{E_{n}-\Delta },
\end{equation}
$E_{1,2}=\pm \sqrt{\left\vert \lambda \right\vert ^{2}+(2\Delta +i\kappa )^{2}/16}$ are the eigenenergies, $t$ is the interaction time, and $\left\vert u_{n}^{r}\right\rangle $ ($n=1,2$) denote the right eigenvectors, which can be written as
\begin{equation}
\left\vert u_{n}^{r}\right\rangle =\frac{\lambda ^{\ast }\left\vert
e,0\right\rangle +\left(E_{n}-\Delta \right)\left\vert g,1\right\rangle }{\sqrt{%
\left\vert \lambda \right\vert ^{2}+\left\vert E_{n}-\Delta \right\vert ^{2}}%
}.
\end{equation}
 The relation between the output state $\left\vert \psi _{NH}\right\rangle$ and $\left\vert u_{n}^{r}\right\rangle$ implies that $\left\vert u_{n}^{r}\right\rangle $ can be extracted from $\left\vert \psi_{NH}\right\rangle $, measured for different interaction times. We obtain $\left\vert u_{n}^{r}\right\rangle $ through the least-squares fitting to the output density matrices measured and post-projected to the single-excitation subspace \cite{Phys.Rev.Lett.131.260201}.

For a certain point in the parameter space of the NH system, the eigenenergies and the corresponding eigenvectors can be written as
\begin{equation}
    E_n=a_n+ i b_n,\ \vert u_n^r\rangle = \sqrt{1-c_n^2}\vert g, 1\rangle + c_n e^{i d_n}\vert e,0\rangle,
\end{equation}
where $a_n$, $b_n$, $c_n$ and $d_n$ are the fitting parameters ($n=1, 2$). To minimize the fitting errors, we define an error function as guidance, which has the form of
\begin{equation}\label{erf}
    {\rm F^{2}_{err}} = {\rm Tr}[\rho \left\vert \psi_{NH}\right\rangle\langle\psi _{NH}\vert]-1.
\end{equation}
Based on this error function, the next step is to seek the fitting parameters that can minimize Eq.~(\ref{erf}) at each moment. In this way, a reliable least-squares fitting to the entire evolution of the density matrix $\rho$ is found with all the parameters.
\begin{figure}
    \centering
    \includegraphics{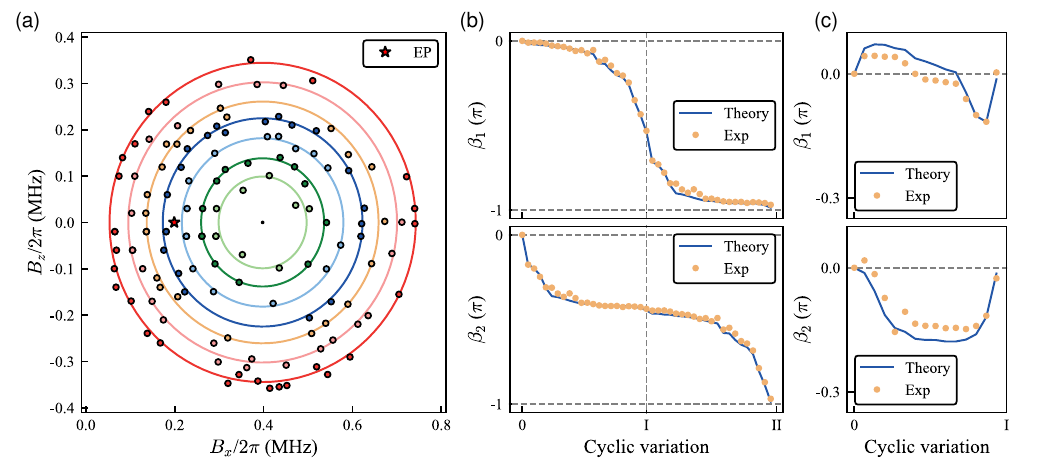}
    \caption{(online) (a) The summation path of the Berry phase on the $B_x$-$B_z$ plane ($B_y=0$) for $B_r/2\pi=0.10$,~$0.14$,~$0.18$,~$0.22$,~$0.26$,~$0.30$,~$0.34$ MHz. The red star marks the exceptional point. The lines denote the theoretical paths and the dots represent the experimental parameters. (b)(c) The corresponding Berry phase after two cycles ($B_r/2\pi=0.26$ MHz) and one cycle ($B_r/2\pi=0.18$ MHz), respectively. The dots represent the experimental data and the solid lines denote the theoretical results for the experimental parameters.}
    \label{figS6_1}
\end{figure}

To demonstrate the validity of this method, we calculate the fidelities of the eigenvectors for the theoretical results along the circular loops, which are centered at ($B_x=\kappa/2$, $B_z$=0) on the $B_x-B_z$ plane, that is
\begin{equation}
    \mathcal{F}_n(\varphi)=\vert\langle u_n^{r'}(\varphi)\vert u_n^r(\varphi)\rangle \vert^2,
\end{equation}
where $\vert u_n^{r'} \rangle$ are the theoretical eigenvectors. In our experiment, we set the radius of the circle $B_r\approx 0.34 \kappa$ and calculate the function with different angles $\varphi$ between the control parameter \textbf{B} and the x-axis. As shown in Fig.~\ref{figS5_2},  the results demonstrate that the eigenstates, extracted from the measured two-qubit output density matrices by our density-matrix post-projecting method, well  agree with the ideal ones, associated with the no-jump evolution trajectories.  With thus-obtained right eigenvectors, the left eigenvectors can be obtained from the biorthonormal condition:

\begin{equation}
\left\langle u_{n}^{l}\right\vert \left. u_{m}^{r}\right\rangle =\delta
_{m,n}.
\end{equation}

\section{Extraction of the Berry phase}
\begin{figure}
    \centering
    \includegraphics{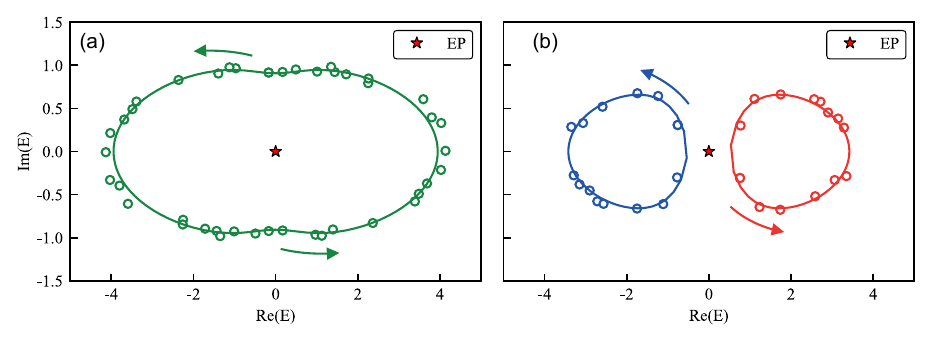}
    \caption{(online) (a) The spectrum along the summation path of the Berry phase across two cycles ($B_r/2\pi=0.26$ MHz).  Due to the eigenvectors exchange, the spectrum's summation path $E_1$ is the same as one of $E_2$, here we only show one of the spectrum. (b) The spectrum along the summation path of the Berry phase across one cycle for $E_1$ (left circle) and $E_2$ (right circle) ($B_r/2\pi=0.18$ MHz), respectively. The red star marks the exceptional point. The dots represent the experimental data and the solid lines denote the theoretical results for the experimental parameters.}
    \label{figS6_2}
\end{figure}

Using the right and left eigenvectors, we can obtain the Berry phase along the path $2{\cal L}$ across the ring twice on the Riemann surface [Fig. 2(a)],
\begin{equation}
\beta _{n}=i%
\displaystyle\oint %
\limits_{2{\cal L}}\left\langle u_{n}^{l}({\bf B})\right\vert \frac{\partial
}{\partial {\bf B}}\left\vert u_{n}^{r}({\bf B})\right\rangle \cdot d{\bf B.}
\end{equation}
In our experiment, we choose circular loops centered at ($B_x = \kappa/2$, $B_z = 0$) on the $B_x$-$B_z$ plane and approximately calculate the Berry phase versus radius $B_r$ through the discretized summation
\begin{equation}
    \beta_{n}(B_r) = i\sum_{p=1}^{2\mathcal{P}}\langle u_{n,p}^{l}(B_r)\vert\frac{\vert u_{n,p+1}^{r}(B_r)\rangle- \vert u_{n,p}^{r}(B_r)\rangle}{\Delta \varphi}\Delta \varphi,
\end{equation}
where $\varphi$ is the angle between the control parameter \textbf{B} and the x-axis, and $\mathcal{P}$ is the number of steps for traversing each cycle. The loop with different radius $B_r$ on the $B_x$-$B_z$ plane is shown in Fig.~\ref{figS6_1}a. Experimentally, the system is excited independently at each step $p$ during stationary-state measurements. The eigenvectors, obtained from these measurements, inevitably carry random phase factors. To track the evolution of the Berry phase in the intermediate steps ($p>1$), we rotate the right eigenvector $\vert u_{n,p+1}^{r}\rangle$ by a phase factor $\phi_{n}^{p+1}=-\arg[\ln \langle u_{n,p+1}^{l} \vert \bar{u}_{n,p}^{r} \rangle]$ for each intermediate step $p$, thus removing the arbitrary phase \cite{Phys.Rev.Lett.127.034301}
\begin{equation}
    \vert \bar{u}_{n,p+1}^{r} \rangle = e^{i\phi_{n}^{p+1}}\vert u_{n,p+1}^{r} \rangle.
\end{equation}
As a result, $\langle \bar{u}_{n,p+1}^{l} \vert \bar{u}_{n,p}^{r} \rangle$ is real and positive, satisfying the condition of parallel transport. For the initial step $p=1$, we set $\vert \bar{u}_{n,p=1}^{r} \rangle = \vert u_{n,p=1}^{r} \rangle$, indicating the initial phase factor is carried along the path $2\mathcal{L}$. This method allows us to track the Berry phase's evolution, shown in Fig.~\ref{figS6_1}b and c. The spectrum along the path is shown in Fig.~\ref{figS6_2}. When $B_r>\kappa/4$, the WER is encircled, and the eigenvectors exchange occurs near $\varphi=\pi$. After the completion of two cycles (Fig.~\ref{figS6_2}a), the eigenvector is restored and obtains a phase difference, approaching $-\pi$, which is the Berry phase. However, when $B_r<\kappa/4$, the WER is not encircled and there is no eigenvectors exchange. So the eigenvector is fully restored in only one complete cycle (Fig.~\ref{figS6_2}b). And the corresponding Berry phase drops to $0$.

\section{Extraction of the Chern number}
The Chern number is calculated on a spherical manifold, which is centered at the origin of the parameter space, i.e., 
\begin{equation}
    \mathcal{C}_n(B_r)=\frac{1}{2\pi}\int_{0}^{2\pi}d\phi\int^{\pi}_{0}d\theta F_{\theta\phi}^n(B_r,\theta,\phi) ,
\end{equation}
\begin{figure}
    \centering
    \includegraphics{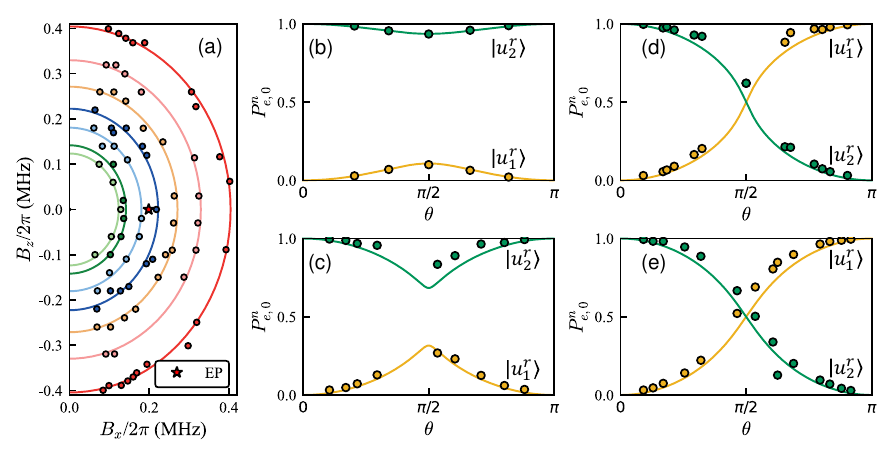}
    \caption{(online) (a) The summation path of the Chern number on the $B_x$-$B_z$ plane ($B_y=0$) for $B_r/2\pi=0.12$,~$0.14$,~$0.18$,~$0.22$,~$0.27$,~$0.32$,~$0.40$ MHz. The red star marks the exceptional point. The lines denote the theoretical path and the dots represent the experimental parameters. Measured populations of $\vert e,0\rangle$ in $\vert u_n\rangle$ ($n=1,2$) versus $\theta$ for $B_r/2\pi=$ $0.12$ MHz (b), $0.18$ MHz (c), $0.22$ MHz (d), $0.27$ MHz (e).}
    \label{figS7}
\end{figure}

where $B_r$ is the radius of the sphere, $\theta$ ($\phi$) is the polar (azimuth) angle, and $F_{\theta\phi}^n\left(B_r,\theta,\phi \right)$  is the Berry curvature. The Berry curvature can be calculated by the Berry connection
\begin{equation}
F_{\theta\phi}^n \left( B_r,\theta,\phi \right)=\partial_\theta A_\phi^n-\partial_\phi A_\theta^n,
\end{equation}
where $ A_\phi^n=i\langle u_n \left(B_r,\theta,\phi \right)\vert\partial_\phi\vert u_n\left(B_r,\theta,\phi \right)\rangle$ and $A_\theta^n=i\langle u_n \left(B_r,\theta,\phi \right)\vert\partial_\theta\vert u_n \left(B_r,\theta,\phi \right)\rangle$.  In our experiment, due to the spherical symmetry, the Berry connection is independent of $\phi$, so that $F_{\theta\phi}^n\left(B_r,\theta \right)=\partial_\theta A_\phi^n \left(B_r,\theta \right)$. The corresponding Chern number is
\begin{equation}
\begin{split}
    \mathcal{C}_n(B_r)=&\frac{1}{2\pi}\int_{0}^{2\pi}d\phi\int^{\pi}_{0}d\theta F_{\theta\phi}^n(B_r,\theta,\phi)\\
    =&\int^{\pi}_{0}d\theta\partial_\theta A_\phi^n(B_r,\theta)\\
    =& P_{e,0}^n(B_r,\pi) - P_{e,0}^n(B_r,0),
\end{split}
\end{equation}
where $P_{e,0}^n(B_r,\theta)$ is the population of $\vert e,0\rangle$ in the eigenvector $\vert u_n^r(B_r,\theta,0)\rangle$ . The paths with different radius $B_r$ we choose to fit the function $P^n_{e,0}$ are shown in Fig.~\ref{figS7}a, and the fitting results $P^n_{e,0}(\theta)$ for different radius $B_r$  are shown in Fig.~\ref{figS7}b-e. Extracting the eigenvector using the method described in Sec.~\ref{S5} is challenging when $\lambda$ is small. Therefore, we calculate the Chern number using $P_{e,0}^n(\theta)$, where $\theta$  is chosen to be as close to $\pi$ and $0$ as possible.
